\documentclass[twocolumn,tighten,times]{aastex63}
\usepackage{xcolor}
\usepackage{apjfonts}

\newcommand{\cajwl}{$Ca_{\rm JWL}$}
\newcommand{\chjwl}{$ch_{\rm JWL}$}
\newcommand{\nhjwl}{$nh_{\rm JWL}$}
\newcommand{\cnjwl}{$cn_{\rm JWL}$}

\newcommand{\nfehhk}{$n$(MP):$n$(MR)}
\newcommand{\nthree}{$n$(P):$n$(I):$n$(E)}

\newcommand{\str}{Str\"omgren}

\newcommand{\vvhbmag}{$-$2.5 mag $\leq$ $V - V_{\rm HB}$ $\leq$ 3.0 mag}
\newcommand{\cnwave}{$\lambda$3883}
\newcommand{\chwave}{$\lambda$4250}
\newcommand{\nhwave}{$\lambda$3360}

\newcommand{\dy}{$\Delta Y$}

\newcommand{\hkjwl}{$hk_{\rm JWL}$}

\newcommand{\fehhk}{[Fe/H]$_{hk}$}
\newcommand{\cfech}{[C/Fe]$_{ch}$}
\newcommand{\nfenh}{[N/Fe]$_{nh}$}

\newcommand{\feh}{[Fe/H]}

\newcommand{\msun}{$M_{\rm \odot}$}

\shorttitle{M92}
\shortauthors{Lee}

\begin{document}

\title{M92 (NGC~6341) is a Metal-Complex Globular Cluster with an Atypical Primordial Population}

\author[0000-0002-2122-3030]{Jae-Woo Lee}
\affiliation{Department of Physics and Astronomy, Sejong University, 209 Neungdong-ro, Gwangjin-Gu, Seoul, 05006, Republic of Korea, 
jaewoolee@sejong.ac.kr, jaewoolee@sejong.edu}

\begin{abstract}
We present a multiple stellar population study of the metal-poor globular cluster (GC) M92 (NGC 6341), which is long known for the substantial metallicity dispersion, using our own photometric system. We find two groups with slightly different mean metallicities, the metal-poor (MP) stars with \fehhk\ = $-$2.412$\pm$0.03,  while the metal-rich (MR) ones with $-$2.282$\pm$0.002. The MP constitutes about 23\% of the total mass with a more central concentration. Our populational tagging based on the \cfech\ and \nfenh\ provides the mean \nthree\ = 32.2:31.6:36.2 ($\pm$2.4), where P, I, and E denote the primordial, intermediate, and extreme populations, respectively. Our populational number ratio is consistent with those of others. However, the MP has a significantly different populational number ratio than the mean value, and the domination of the primordial population in the MP is consistent with observations of Galactic GCs that less massive GCs contain larger fractions of the primordial population. Structural and constituent differences between the MP and MR may indicate that M92 is a merger remnant in a dwarf galaxy environment, consistent with recent suggestions that M92 is a GC in a dwarf galaxy or a remnant nucleus of the progenitor galaxy.
Discrepancy between our method and those widely used for the HST photometry exists in the primordial population. Significant magnesium and oxygen depletions of $-$0.8 and $-$0.3 dex, respectively, and helium enhancement of \dy\ $\gtrsim$ 0.03 are required to explain the presence of this abnormal primordial group. No clear explanation is available with limited information of detailed elemental abundances.
\end{abstract}


\section{INTRODUCTION}
A decade-long puzzle of light elemental abundance variations in Galactic globular clusters (GCs) is now explained with multiple stellar populations (MSPs), where the second generation (SG) of stars formed out of interstellar media polluted by the first generation (FG) of stars \citep[e.g.,][]{dercole08, bekki19}. During the past decade, a great deal of observational evidence and theoretical progress have been accumulated to explain the GC MSPs not only in our Milky Way but also in nearby galaxies \citep[e.g.,][]{milone17, bastian18}. Except for a small number of some special GCs, such as $\omega$ Cen, intracluster inhomogeneity in metal abundances is often neglected.
However, the conventional assumption of the homogeneity in iron abundances of normal GCs has been challenged recently \citep[e.g.,][]{lee21a, lee22, legnardi22, lardo22}.

Spectroscopic study of individual chemical elements of GC stars is believed to provide an ultimate verdict on elemental abundance variations.
However, the detection of a small-scale dispersion in metal abundances in GCs can be a challenging task with seemingly straightforward 1D local thermodynamic equilibrium analysis of high-resolution spectra. Disagreement in detailed results among different research groups can often be witnessed \citep[e.g.,][for 47~Tuc]{lee22}. Alternatively, high-precision narrowband photometry of GC stars can provide a wonderful opportunity to study underlying stellar populations. One of the main characteristics of GC MSPs is the C--N anticorrelation, due to a consequence of the CN cycle. The carbon and nitrogen abundance variations can be photometrically determined through very strong absorption band features of NH, CN, and CH molecules.
In this regard, we have developed a series of narrowband photometric systems that can measure metallicity and carbon and nitrogen abundances in GC red giant branch (RGB) stars \citep[e.g.,][]{lee15, lee22, lee21a}.

M92 has been paid little attention until recently, in spite of intriguing results on the metallicity spread more than two decades ago. By employing high-resolution spectroscopy, \citet[][for subgiants]{king98} and \citet[][for RGBs]{langer98} reported substantial metallicity spread by about 0.15--0.2 dex in M92. More recent results of \citet{roederer11} showed the existence of metallicity spread and the variations in the heavy neutron-capture elemental abundances as well.
In a similar context, M92 is known to show an extended FG population in its chromosome map \citep{milone17}. In our previous work for M3, we successfully showed that this extended FG is due to metallicity spread \citep{lee21a}, confirmed later by \citet{lardo22} and \citet{legnardi22} for other GCs, including M92.

\begin{figure*}
\epsscale{1.1}
\figurenum{1}
\plotone{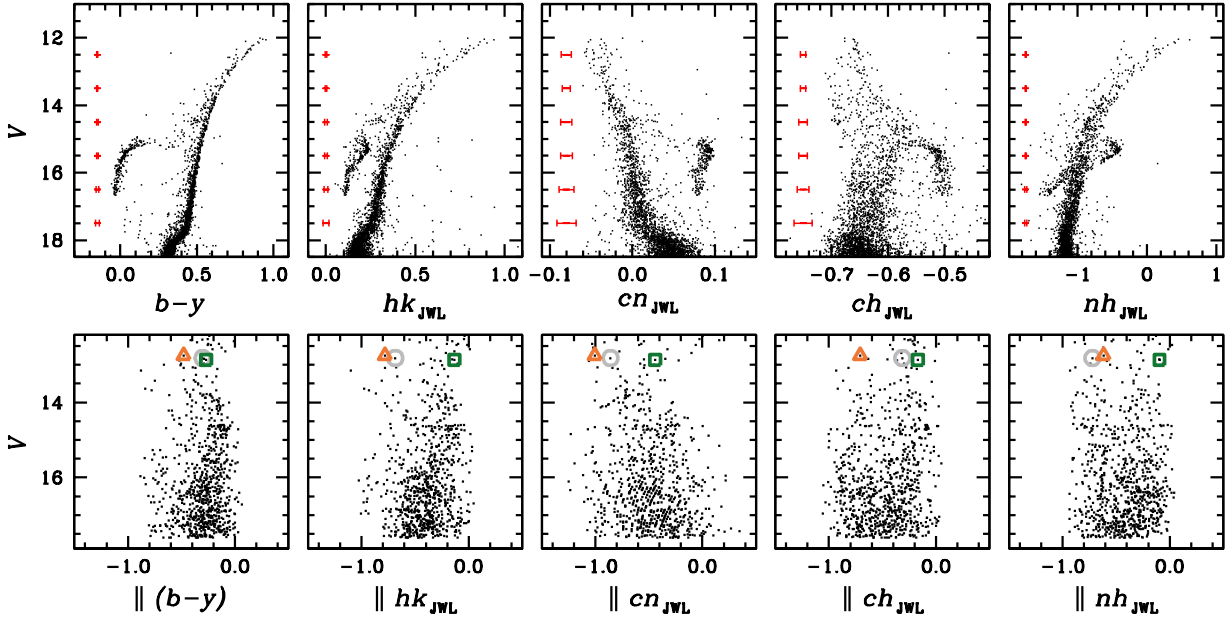}
\caption{(Top) CMDs for proper motion membership stars in M92. We also show the mean measurement uncertainties at a given $V$ magnitude bin with red error bars.
(Bottom) Parallelized CMDs of RGB stars with \vvhbmag. In each panel, we show stars V-45 (open gray circles), XII-8 (open orange triangles), and XI-19 (open green squares).
}\label{fig:cmd}
\end{figure*}

Recently, a stellar stream related to M92 has been discovered \citep{sollima20, thomas20, martin22}. At face value, the stellar stream is dynamically young, and its estimated mass could be as large as $\sim$ 10\% of the total cluster's mass. \citet{thomas20} drew an interesting suggestion that M92 could be a GC that was brought into the Milky Way by a dwarf galaxy or a remnant nucleus of the progenitor galaxy.

In this letter, we present a photometric metallicity, [C/Fe], and [N/Fe] study for the metal-poor GC M92, to shed more light on the complexity of the cluster.

\section{OBSERVATIONS AND DATA REDUCTION}\label{s:reduction}
Observations for M92 were carried out in 15 nights, five of which  were photometric, in five runs from 2017 April to 2019 July using the KPNO 0.9 m telescope.
In 2017 and 2018, we obtained photometric data using the Half Degree Imager (HDI), which is equipped with an e2V 4 $\times$ 4k CCD chip, mounted on the the KPNO 0.9m telescope.
The HDI provides a field of view (FOV) of 30\arcmin $\times$ 30\arcmin. 
Additional photometric data for M92 using our new $JWL34$ filter were collected in 2019 June 27 through July 5.
Due to an electronic problem with the HDI in 2019, we used the S2KB CCD camera, which is equipped with a 2 $\times$ 2k CCD chip and provides an FOV of 21\arcmin $\times$ 21\arcmin.

The total integration times of \str\ $y$, $b$, \cajwl, $JWL39$, $JWL43$, and $JWL34$ for the M92 science field were 3160 s, 6930 s, 20700 s, 11400 s, 11600 s, 21800 s, respectively.

The raw data handling was described in detail in our previous works \citep{lee15, lee17}. The photometry of M92 and standard stars were analyzed using DAOPHOTII, DAOGROW, ALLSTAR and ALLFRAME, and COLLECT-CCDAVE-NEWTRIAL packages \citep{pbs87, pbs94}.
Finally, we derived the astrometric solutions for individual stars using the Gaia Early Data Release 3 \citep[EDR3;][]{gaiaedr3} and the IRAF IMCOORS package.

In order to select M92 member stars, we made use of the proper-motion study from the Gaia EDR3. We derived the mean proper-motion values using iterative sigma-clipping calculations, finding that, in units of milliarcsecond per year, ($\mu_{\rm RA}\times\cos\delta$, $\mu_{\rm decl.}$) = ($-$4.892, 0.830) with standard deviations along the major axis of the ellipse of 0.830 mas~yr$^{-1}$ and along the minor axis of 0.826 mas~yr$^{-1}$. We considered that stars within 3$\sigma$ from the mean values to be M92 proper-motion member stars.

Due to rather large plate scales of CCD chips that we used, about 0.43 and 0.62 \arcsec/pixel for the HDI and S2KB, respectively, combined with poor detection completeness of Gaia in the crowded field, our result suffers from an incomplete detection of stars in the central part of the cluster. By comparing with the HST photometric data by \citet{uvlegacy}, the detection probability is $\sim$ 15\% in the central 10\arcsec\ and reaches 100\% at the radial distance of $\sim$ 110\arcsec.

We note that incomplete detection of stars in the central part of M92 does not likely affect our results presented here. We performed Monte Carlo simulations using our presumed populational number ratios and the detection probability mentioned above with varying full-width of half maximums (FWHMs) of the distribution of the FG with respect to that of
the SG. Our simulations suggest that incomplete detection does not greatly affect the initial populational number ratios and the central concentration ranking between the two groups of stars statistically significantly.
\begin{figure}
\epsscale{1.}
\figurenum{2}
\plotone{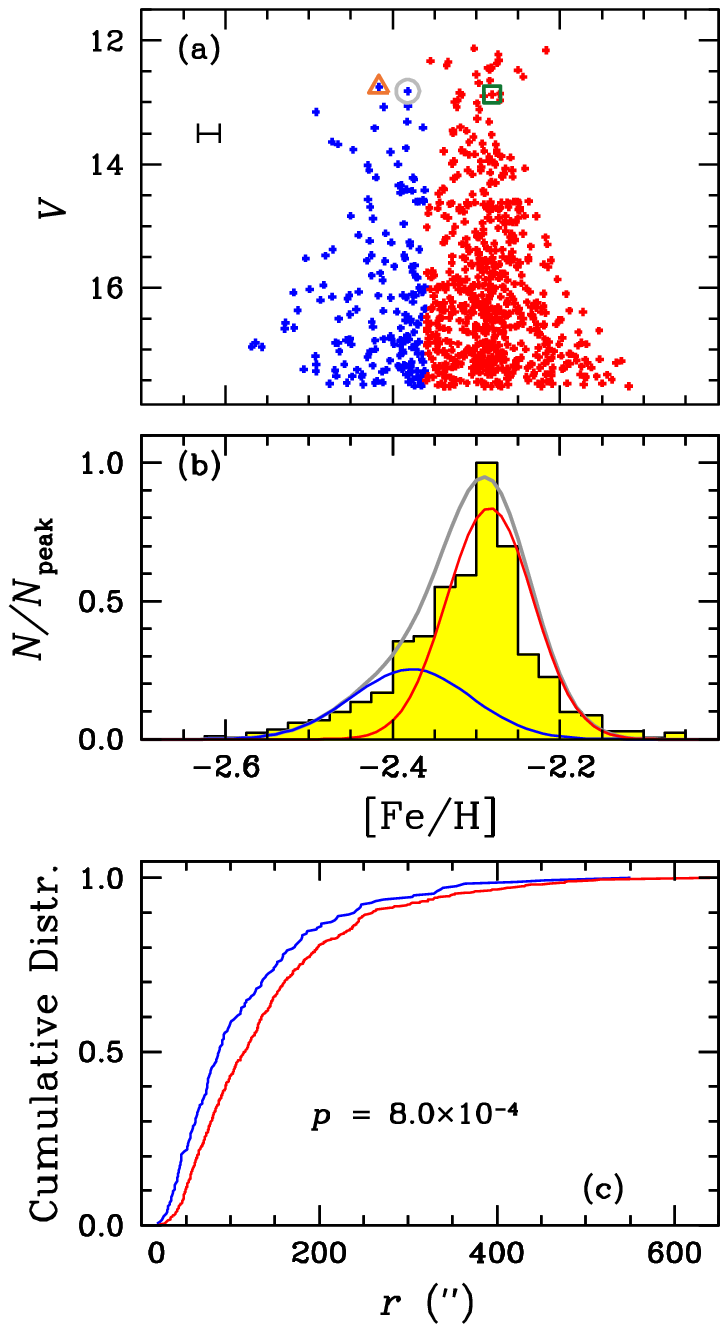}
\caption{(a) Metallicity distributions of individual RGB stars with \vvhbmag. The blue and red colors denote the metal-poor and metal-rich populations in M92.
The measurement uncertainties ($\pm$1$\sigma$) are shown.
We show stars V-45 (open gray circle), XII-8 (open orange triangle), and XI-19 (open green square) studied by \citet{langer98}.
(b) Generalized histograms returned from our EM estimator for each population along with the normalized observed histogram. The blue and red lines are for the MP and MR populations, while the gray line is for the all RGB stars.
(c) Cumulative distributions. The MP population is more centrally concentrated than the MR population is.}\label{fig:feh}
\end{figure}

\section{Photometric Indices and Color--Magnitude Diagrams}
Throughout this work, we will use our own photometric indices \citep[see also][]{lee19, lee21a, lee22}, defined as
\begin{eqnarray}
{{hk}}_{\mathrm{JWL}} &=& ({\mathrm{Ca}}_{\mathrm{JWL}}-b)-(b-y),\\
{{cn}}_{\mathrm{JWL}} &=& {JWL}39-{\mathrm{Ca}}_{\mathrm{JWL}},\\
{{ch}}_{\mathrm{JWL}} &=& ({JWL}43-b)-(b-y),\\
{{nh}}_{\mathrm{JWL}} &=& ({JWL}34-b)-(b-y),
\end{eqnarray}
The \hkjwl\ index is a good photometric measure of metallicity \citep[e.g.,][]{att91, lee09, lee15}, assuming a constant [Ca/Fe] among GC membership stars \citep{carney96}. The \nhjwl, \cnjwl, and \chjwl\ indices are measures of NH absorption band at \nhwave, CN at \cnwave, and CH at \chwave\ \AA, respectively.

In Figure~\ref{fig:cmd}, we show our color--magnitude diagrams (CMDs) of M92 using our color indices. In order to remove the luminosity effect in RGB stars in our interest with \vvhbmag, the RGB sequences in the individual color indices were parallelized using the following relation \citep[also see][]{lee19, lee22, lee21a}
\begin{equation}
\parallel \mathrm{CI}(x)\equiv \displaystyle \frac{\mathrm{CI}(x)-{\mathrm{CI}}_{\mathrm{red}}}{{\mathrm{CI}}_{\mathrm{red}}-{\mathrm{CI}}_{\mathrm{blue}}},\label{eq:pl}
\end{equation}
where, CI$(x)$ is the color index of the individual stars and CI$_{\rm red}$, CI$_{\rm blue}$ are color indices for the fiducials of the red and the blue sequences of individual color indices, respectively.
We derived fourth-order polynomial fits for individual color indices except for the \chjwl. Note that our \chjwl\ versus $V$ CMD shows a break in the slope of RGB sequence at $V$ $\sim$ 14.6 mag due to internal mixing accompanied by CN cycle in the core that induces the reduction of the surface carbon abundance of bright RGB stars. Therefore, we split our fitting procedures at $V$ = 14.6 mag, and we combined them. We show our parallelized CMDs in the bottom panels of Figure~\ref{fig:cmd}.
We would like to point out two interesting aspects:
\begin{itemize}\setlength\itemsep{0em}
\item Unlike our previous investigations of more metal-rich GCs \citep[e.g.,][]{lee17, lee18, lee22, lee21a}, our \cnjwl\ index, which is a measure of the CN band strength at \cnwave, does not clearly show a bimodal distribution in M92 \citep[see also][]{smolinski11}. Note that the width of the \cnjwl\ sequence of the RGB stars is comparable to the measurement uncertainties. This is due to the double-metal nature of the CN molecule, whose absorption strengths rapidly decline with decreasing metallicity \citep[e.g.,][]{sneden74}.
Due to a similar reason, $(U-B)-(B-I)$, which makes use of the CN red system, is not useful to classify the MSPs in the metal-poor GCs.
\item The widths of observed RGB sequences in other photometric indices, such as $(b-y)$, \hkjwl, \chjwl, and \nhjwl, are much larger than the measurement uncertainties, indicating that there exist variations in relevant elemental abundances, such as metallicity, carbon, and nitrogen abundances \citep{lee21a, lee22}.
\end{itemize}

\begin{figure}
\epsscale{1.15}
\figurenum{3}
\plotone{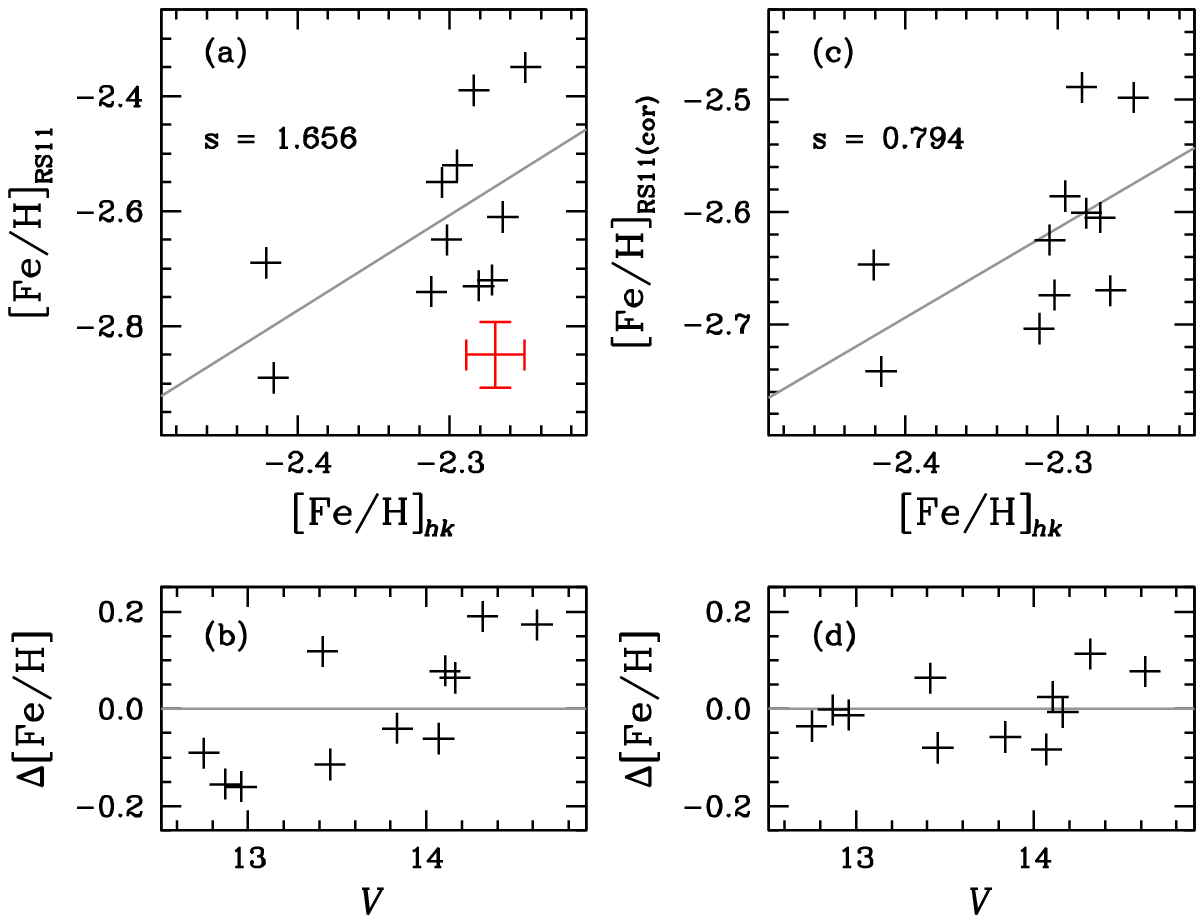}
\caption{
A comparison of our photometric metallicity \fehhk\ to that of high-resolution spectroscopic study by \citet{roederer11}.
(a) A plot of our photometric \feh\ vs. spectroscopic \feh\ by \citet{roederer11}, showing a good correlation with \feh$_{\rm RS11}$ $\propto$ 1.656$\times$\fehhk. The error bar in the bottom right corner denotes the typical measurement errors in both studies.
(b) Residuals around the fitted line against the visual magnitude, showing a substantial gradient.
(c) A plot of our photometric \feh\ vs. magnitude-gradient-corrected \feh\ by \citet{roederer11}, showing \feh$_{\rm RS11,c}$ $\propto$ 0.794$\times$\fehhk.
(d) Residuals around the fitted line against the visual magnitude using (c).
}\label{fig:2rs}
\end{figure}

\begin{figure}
\epsscale{1.1}
\figurenum{4}
\plotone{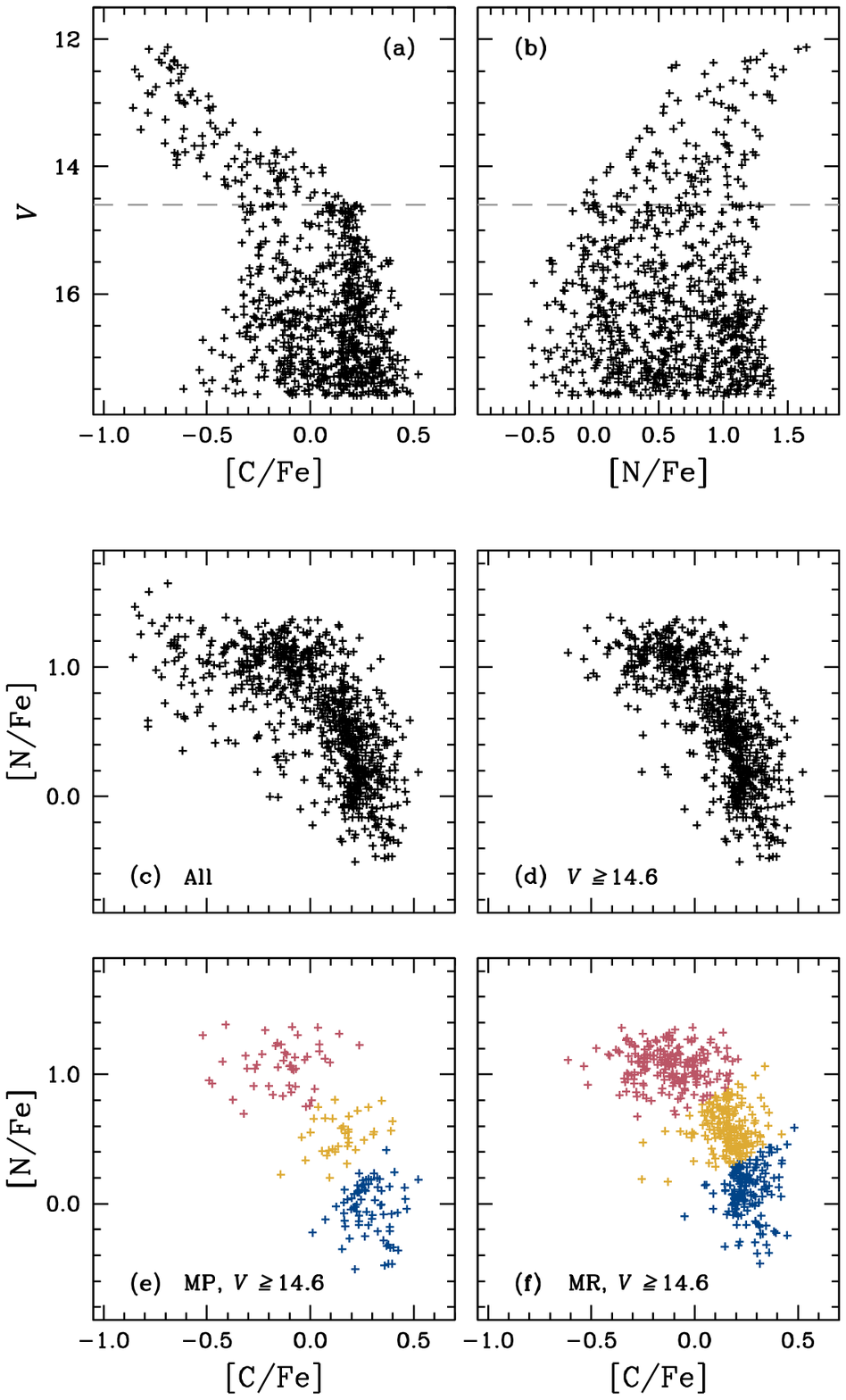}
\caption{
(a) The \cfech\ vs. $V$ magnitude of M92 RGB stars with \vvhbmag. The gray dashed line indicates the $V$ magnitude of the RGBB ($\approx$ 14.6 mag).
(b) Same as (a) but for the \nfenh\ vs. $V$ magnitude.
(c) The \cfech\ vs. \nfenh\ of RGB stars with \vvhbmag.
(d) The \cfech\ vs. \nfenh\ of RGB stars fainter than the RGBB, $V \geq$ 14.6 mag.
(e) The \cfech\ vs. \nfenh\ of MP RGB stars fainter than the RGBB. The blue, gold, and red crosses denote the primordial, intermediate, and extreme subpopulations, respectively.
(f) The \cfech\ vs. \nfenh\ of MR RGB stars fainter than the RGBB. The colors are the same as (e).
}\label{fig:CN}
\end{figure}


\begin{deluxetable}{lcr}[t]
\tablenum{1}
\tablecaption{Mean [Fe/H] values.\label{tab:feh}}
\tablewidth{0pc}
\tablehead{
\multicolumn{1}{c}{} &
\multicolumn{1}{c}{[Fe/H]} &
\multicolumn{1}{c}{frac. (\%)}
}
\startdata
All      & $-$2.311 $\pm$ 0.002 ($\sigma$ = 0.071) & 100.0 \\
MP       & $-$2.412 $\pm$ 0.003 ($\sigma$ = 0.046) &  22.5 \\
MR       & $-$2.282 $\pm$ 0.002 ($\sigma$ = 0.046) &  77.5 \\
\enddata
\end{deluxetable}

\section{Metallicity, Carbon, and Nitrogen Abundances}
Our \hkjwl\ index is an excellent measure of the \ion{Ca}{2} H and K lines \citep[e.g., see][]{att91, lee09, lee15, lee22, lee21a}.
In our previous studies, we obtained the [Fe/H] values of individual RGB stars in M3, M5, and 47~Tuc using our \hkjwl\ \citep{lee21a, lee21b, lee22}. Here, we derive metallicity of individual RGB stars in M92.

In order to derive photometric metallicity, we calculated model atmospheres using ATLAS12 \citep{kurucz11} and synthetic spectra using the latest version of MOOGSCAT that employs a nonlocal thermodynamic equilibrium (NLTE) treatment and improved Rayleigh scattering \citep{moog, moogscat}.
We retrieved the model isochrones for [Fe/H] = $-$2.1, $-$2.3, $-$2.5, $Y$ = 0.247, 0.275, and 0.300 with [$\alpha$/Fe] = +0.4 dex, and the age of 12.5 Gyr from a Bag of Stellar Tracks and Isochrones \citep[BaSTI;][]{basti21}.
We adopted different CNO abundances, [C/Fe] = ($-$0.6, $\Delta$[C/Fe] = 0.2 ,0.6), [N/Fe] = ($-$0.8, $\Delta$[N/Fe] = 0.4 , 1.6), and [O/Fe] = (0.1, 0.3, 0.5) for each model grid. Note that our presumed CNO abundances do not affect our photometric metallicity \citep{lee22}. We constructed 97 model atmospheres and synthetic spectra for each chemical composition from the lower main sequences to the tip of RGB sequences.
For metal-poor stars, the negative hydrogen ion's bound-free (H$^-_{\rm bf}$) absorption dominates continuum opacities in the optical and infrared wavelength domains.
However, in the blue and UV regions, Rayleigh scattering from neutral hydrogen  (RSNH) atoms becomes the dominant source of continuum opacity, especially in cool RGB atmospheres \citep[e.g., see][]{suntzeff80,moogscat,lee21a}.
The latest version of MOOGSCAT \citep{moogscat} takes proper care of RSNH with a NLTE treatment of the source function, which is important to calculate continuum opacities for our short-wavelength indices such as our $JWL34$ (i.e., \nhjwl), due to a $\lambda^{-4}$ dependency of the RSNH cross section. We emphasize that the synthetic \nhjwl\ indices produced using MOOGSCAT satisfactorily reproduce our observations, while those using SYNTHE do not. On the other hand, other color indices, such as \hkjwl\ and \chjwl, are not significantly affected by the choice of codes.

Individual synthetic spectra were convolved with our filter transmission functions to be converted to our photometric system, and the bolometric corrections for individual colors were calculated using the same  method described by \citet{girardi02}.

The photometric metallicity of individual RGB stars can be calculated using the following relation \citep[also see Appendices of][]{lee21a}
\begin{eqnarray}
{\rm [Fe/H]}_{hk} &\approx& f_1({{hk}}_{\mathrm{JWL}},~ M_V).\label{eq:FeH}
\end{eqnarray}
We obtained the mean \fehhk\ = $-$2.311 $\pm$ 0.002 dex ($\sigma$ = 0.071) and we show our results in Figure~\ref{fig:feh}.
Our mean photometric metallicity of M92 is in excellent agreement with previous results by others. For example, \citet{sneden00} obtained \feh\ = $-$2.341 $\pm$ 0.007 ($\sigma$ = 0.039) from 32 stars in M92. Note that our photometric \fehhk\ has a large standard deviation, $\sigma$ = 0.071, due to the double metallicity distributions of M92 as discussed below, in sharp contrast to a small value of \citet{sneden00}, $\sigma$ = 0.039.

\citet{roederer11} performed a differential spectroscopic study of M92 giant stars. They argued that M92 shows inhomogeneity in neutron-capture elemental abundances but the iron peak elemental abundances are homogeneous at the level up to 0.16 dex. Note that the \feh\ values of \citet{roederer11}, with a mean of \feh\ = $-$2.70 $\pm$ 0.03 ($\sigma$ = 0.14), are lower than those of previous studies but with a large dispersion. We compare our photometric \fehhk\ with that of \citet{roederer11} and  show our results in Figure~\ref{fig:2rs}. The figure suggests that our photometric \fehhk\ is in good correlation with the spectroscopic \feh\ by \citet{roederer11} although the slope of the correlation returned from a linear regression is somewhat steep, \feh$_{\rm RS11}$ $\propto$ 1.656$\times$\fehhk.

As \citet{roederer11} noted, their \feh\ measurements show a temperature gradient, which can also be seen in our plot of residuals around the fitted lines against the visual magnitude in Figure~\ref{fig:2rs}(b). On the other hand, our photometric \fehhk\ does not show any gradient against the visual magnitude as shown in Figure~\ref{fig:feh} (a). We attempt to remove the temperature gradient in the \feh\ measurements by \citet{roederer11} by deriving a linear regression in Figure~\ref{fig:2rs}(b). We show our plot of magnitude-gradient-corrected \feh\ by \citet{roederer11} against our photometric \fehhk\ in Figure~\ref{fig:2rs}(c), showing a slope more close to unity, \feh$_{\rm RS11,c}$ $\propto$ 0.794$\times$\fehhk. This result indicates that our photometric \fehhk\ should be a good measure of \feh\ for M92 RGB stars \citep[see also Figure~8 of][]{lee21a}.

In Figure~\ref{fig:feh}, we show three RGB stars studied by \citet{langer98}, who argued heavy elemental abundance spread in M92, in the sense that XI-19 is 0.18 $\pm$ 0.01 dex more metal-rich than V-45 and XII-8. We obtained photometric metallicities of \fehhk\ = $-$2.38 (V-45), $-$2.42 (XII-8), and $-$2.28 dex (XI-19). Although the difference in our photometric metallicity between two groups of stars, 0.12 $\pm$ 0.02, is slightly smaller than that argued by \citet{langer98}, our result confirmed that XI-19 is truly more metal-rich than V-45 and XII-8.

The metallicity distribution of M92 exhibits asymmetric metallicity distributions with long metal-poor tails similar to those that can be seen in M3 and 47~Tuc \citep{lee21a, lee22}. Assuming a bimodal \fehhk\ distribution of M92 RGB population, we applied an expectation-maximization (EM) algorithm for a two-component Gaussian mixture model on our \fehhk\ distribution. Stars with $P$(\fehhk$|x_i) \geq$ 0.5 from the EM estimator correspond to the metal-poor (MP) population, where $x_i$ denotes the individual RGB stars, while those with $P$(\fehhk$|x_i)$ $<$ 0.5  correspond to the metal-rich (MR) population. We obtained the populational number ratio of \nfehhk\ = 22.5:77.5 ($\pm$1.8). We show our result in Figure~\ref{fig:feh}(b) and Table~\ref{tab:feh}. The MP population is about 0.13 dex more metal-poor than the MR population in M92. In the figure, we also show the cumulative radial distributions, finding that the MP population is more centrally concentrated than the MR. We performed a Kolmogorov--Smirnov test and obtained a very small $p$ value, 7.92$\times10^{-4}$, suggesting that both populations are not drawn from the same parent population.

\begin{figure*}
\epsscale{1.2}
\figurenum{5}
\plotone{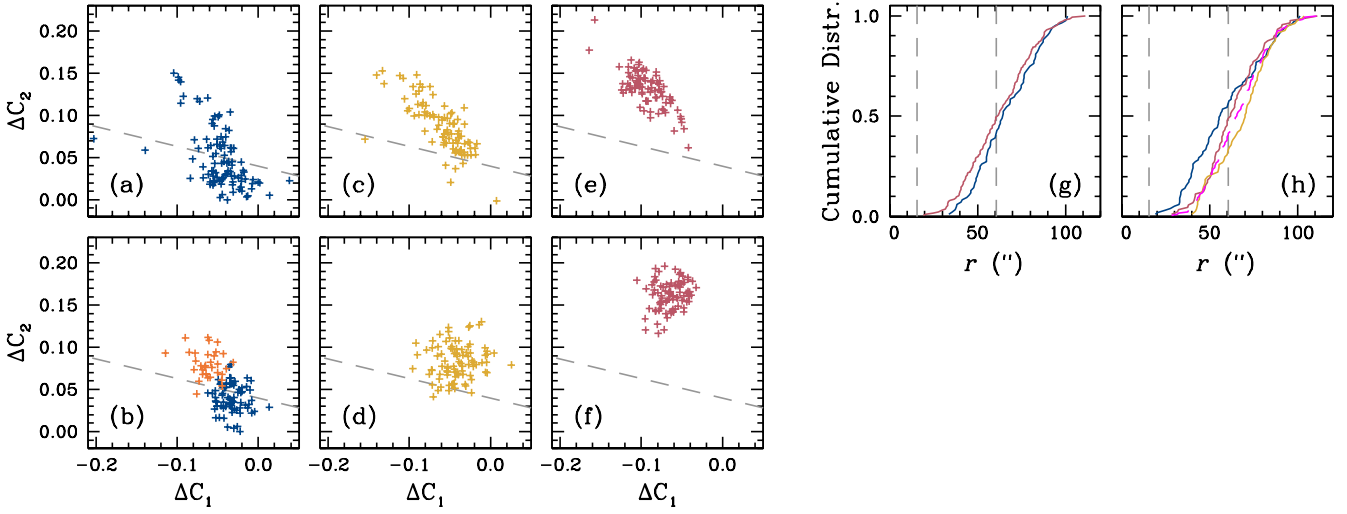}
\caption{
(a) A chromosome map of the primordial population of our study. The gray dashed line shows the border line between the FG and SG populations by \citet{milone17}.
(b) The simulated chromosome map for our primordial population. The orange plus signs (P2) are for ($\Delta Y$, $\Delta$[O/Fe], $\Delta$[Mg/Fe]) = (+0.028, $-$0.30, $-$0.80) with respect to the P1 (blue plus signs).
(c) Same as (a) but for the intermediate population.
(d) The simulated chromosome map for our intermediate population, with ($\Delta Y$, $\Delta$[C/Fe], $\Delta$[N/Fe], $\Delta$[O/Fe]) = (+0.028, $-$0.30, +0.60, $-$0.20) with respect to the P1.
(e) Same as (a) but for the extreme population.
(f) The simulated chromosome map for our intermediate population, with ($\Delta Y$, $\Delta$[C/Fe], $\Delta$[N/Fe], $\Delta$[O/Fe]) = (+0.028, $-$0.60, +1.20, $-$0.40) with respect to the P1.
(g) CRDs of the FG (the blue solid line) and SG (the red solid line) classified by using the border line by \citet{milone17}.
(h) CRDs of the primordial (the blue solid line), intermediate (the gold solid line), and extreme (the red solid line) populations. The magenta dashed line is for the enhanced population.
}\label{fig:hst}
\end{figure*}

We also derive the photometric \cfech\ and \nfenh\ from our \chjwl\ and \nhjwl\ indices using the following relations \citep{lee21b, lee21a},
\begin{eqnarray}
{\rm [C/Fe]}_{ch} &\approx& f_2({{ch}}_{\mathrm{JWL}},~{\mathrm{[Fe/H]}_{hk}}, ~ M_V), \\
{\rm [N/Fe]}_{nh} &\approx& f_3({{nh}}_{\mathrm{JWL}},~{\mathrm{[Fe/H]}_{hk}}, ~ M_V).
\end{eqnarray}
In Figure~\ref{fig:CN}(a--b), we show \cfech\ and \nfenh\ against the $V$ magnitude.
Our results clearly show that the carbon abundance decreases and nitrogen abundance increases in RGB stars brighter than the RGB bump (RGBB, $V$ $\approx$ 14.6 mag for M92). In Figure~\ref{fig:CN}(c--d), we show plots of \cfech\ versus \nfenh, showing a carbon--nitrogen anticorrelation, a natural consequence of the CN-cycle hydrogen burning.

We classify the underlying M92 subpopulations based on our \cfech\ and \nfenh.
We calculate the histogram of the [C/N] distribution of individual stars with $V$ $\geq$ 14.6 mag, finding three distinctive peaks. Our examination of the density plot of the \cfech\ versus \nfenh\ also reveals three peaks, which supports our finding of three subpopulations.
We applied an EM algorithm for a three-component Gaussian mixture model on our [C/N] distribution, finding the number ratio of \nthree\ = 32.2:31.6:36.2 ($\pm$2.4), where P, I, and E stand for the primordial, intermediate, and extreme populations \citep[e.g.,][]{lee21b}. We show individual populations in Figure~\ref{fig:CN}(e--f). Note that our populational number ratio of the primordial population is in good agreement with that of \citet{milone17}, $n$(FG):$n$(SG) = 30.4:69.6 ($\pm$1.5).

We note that the subpopulational number ratios are significantly different between the MP and MR populations, as shown in Figure~\ref{fig:CN}(e--f). We obtained \nthree\ = 45.9:22.3:31.8 ($\pm$4.2) for the MP, while we obtained 28.3:34.3:37.4 ($\pm$2.9) for the MR. The predominance of the primordial subpopulation in the MP appears to be consistent with the idea that less massive GCs tend to contain higher fraction of the primordial population \citep[e.g.][]{milone20} if the MP and MR are independent GC systems when they formed.

\section{An Atypical Primordial Population}
Finally, we made a comparison of our study with that of \citet{milone17}. In Figure~\ref{fig:hst}, we show plots of $\Delta$C$_1$ (= $\Delta_{\rm F275W,F814W}$) versus $\Delta$C$_2$(= $\Delta_C\ _{\rm F275W,F336W,F438W}$), i.e., a chromosome map, for common stars fainter than the RGBB between our study and that of \citet{uvlegacy}.
As shown in the figure, it is interesting to note that our primordial population is distributed both in the FG and SG domains defined by \citet{milone17}. Note that the distributions of our intermediate and extreme populations, which contain nitrogen-enhanced populations, on the chromosome map are consistent with those in other GCs. Therefore, we argue that the distribution of our primordial population on the chromosome map is not in great error.
We argue that our definition of the GC MSPs, based on metallicity, carbon, and nitrogen abundances, is more straightforward than that defined on the chromosome map, which includes multiple sources with potential degeneracy \citep[e.g., see Figure~8][]{milone18}.
In the following discussion, the P1 denotes the primordial population below the border line (i.e., lower $\Delta$C$_2$ values) and the P2 is for those above the border line. We examined various aspects between the two groups of stars, and we found the following.
\begin{itemize}\setlength\itemsep{0em}
\item The P1 and P2 have the same mean \fehhk, \cfech, and \nfenh.
\item In spite of the small numbers of bright RGB stars, the RGBB $V$ magnitude of the P2 appears to be brighter than that of the P1, suggesting that the P2 is more helium enhanced. This is also consistent with the P2 distribution on the chromosome map.
\item The P2 appears to be more centrally concentrated than the P1.
\end{itemize}

In order to understand the physical differences between the P1 and P2, we calculated synthetic spectra with various chemical compositions and performed evolutionary population synthesis. We show our results in Figure~\ref{fig:hst}. With fixed metallicity, carbon, and nitrogen abundances between the two groups of stars, the P2 requires significant depletion in the magnesium abundance by about $-$0.8 dex or more, along with an oxygen depletion of about $-$0.3 dex.\footnote{Magnesium affects the HST F275W passband through \ion{Mg}{2} h and k lines, while oxygen mainly affects the HST F275W and F336W through OH molecules.} In addition, the helium is enhanced by $\Delta Y$ $\gtrsim$ 0.03, from the difference in the RGBB $V$ magnitudes.

\section{SUMMARY}
We presented a narrowband photometry of M92 using our own photometric system. Our populational number ratio for M92 based on the carbon and nitrogen abundances,\nthree\ = 32.2:31.6:36.2 ($\pm$2.4), is consistent with those of others, $n$(FG):$n$(SG) = 30:70 \citep[e.g.][]{milone17}, with our intermediate and extreme populations being the SG.

We argued that M92 contains two metallicity groups. The MP constitutes about 23\% of the total mass with a more central concentration, similar to what we found in M3 and 47~Tuc \citep{lee21a, lee22}.
We obtained \nthree\ = 45.9:22.3:31.8 ($\pm$4.2) for the MP, while 28.3:34.3:37.4 ($\pm$2.9) for the MR. Our results are in sharp contrast to those of inferred metal inhomogeneity solely in the FG using the chromosome map by others \citep{legnardi22, lardo22}. The large fraction of the primordial population in the MP is consistent with the fact that the primordial population is more frequent in less massive GCs \citep[e.g.,][]{milone20} if the MP and MR are independent GC systems. At the current mass of M92 \citep{baumgardt18}, the total mass of the MP subpopulations is $\sim$ 6$\times$10$^4$\msun, comparable to those of less massive GCs observed in our Galaxy.

The more centrally concentrated nature of the MP cannot be explained with the diffusion process. The evolutionary stellar mass of the MP population is smaller than the MR. Therefore, the MP population tends to exhibit more spatially extended distribution than the MR during the dynamical evolution of M92 if the initial radial distributions of both the MP and MR were similar. At the same time, if the MP is the precursor of the MR in an in situ formation, the MR would be more centrally located in the cluster \citep[e.g.,][]{calura19}, without mentioning that the current total MP mass is too small to account the metallicity enhancement of the MR. Therefore, one of the simple explanations of M92 with different subpopulational number ratios and cumulative radial distributions between the MP and MR would be a merger of two GCs, most likely in a dwarf galaxy environment,  where the merger rates are much higher than our Galaxy \citep[e.g.,][]{thurl02}, as we proposed for M22, M3, and 47~Tuc \citep{lee15, lee22, lee21a}. The conclusion drawn by \citet{thomas20}, that M92 could be a GC that was brought into the Milky Way by a dwarf galaxy or a remnant nucleus of the progenitor galaxy, may support our suggestion.

By comparing the HST photometric data, we found a discrepancy between our method and those widely used for the HST photometry. Our primordial RGB stars defined in our \cfech\ and \nfenh\ domain are distributed both in the FG and SG domain of the chromosome map. Our simulations suggested that the P2 has depleted oxygen and magnesium abundances by $-$0.3 and $-$0.8 dex, respectively, and enhanced helium abundance by \dy\ $\gtrsim$ 0.03 with respect to the P1, while both have the similar metallicity, carbon and nitrogen abundances. The elemental abundance difference between the two cannot be due to an internal mixing \citep[e.g.,][]{lee10}, since our RGB stars are fainter than the RGBB.
With our limited information of chemical abundances of other elements, such as aluminum, no clear explanation is available for the P2.
Future study in this regard would be very desirable to reveal the formation history of M92.

\acknowledgements
{J.-W.L.\ acknowledges financial support from the Basic Science Research Program (grant no.\ 2019R1A2C2086290) through the National Research Foundation of Korea (NRF) and from the faculty research fund of Sejong University in 2022. We thank Jennifer Sobeck and Chris Sneden for providing the latest version of MOOGSCAT.
We also thank an anonymous referee for useful comments and a careful review of the paper.}

\facilities{WIYN:0.9m (HDI, S2KB), Gaia}


\begin{thebibliography}{}

\bibitem[Anthony-Twarog et al.(1991)]{att91}
Anthony-Twarog, B.~J., Laird, J.~N., Payne, D., \& Twarog, B.~A.\ 1991, \aj, 101, 1902

\bibitem[Bastian \& Lardo(2018)]{bastian18} Bastian, N., \& Lardo, C.\ 2018, \araa, 56, 3

\bibitem[Baumgardt \& Hilker(2018)]{baumgardt18}
Baumgardt, H., \& Hilker, M.\ 2018, \mnras, 478, 1520

\bibitem[Bekki(2019)]{bekki19} Bekki, K.\ 2019, \aap, 622, 53

\bibitem[Calura et al.(2019)]{calura19}
Calura, F., D'Ercole, A., Vesperini, E., Vanzella, E., \& Sollima, A.\ 2019, \mnras, 489, 3269

\bibitem[Carney(1996)]{carney96} Carney, B.~W.\ 1996, \pasp, 108, 900

\bibitem[D'ercole et al.(2008)]{dercole08}
D'Ercole, A., Vesperini, E., D'Antona, F., McMillan, S.\ L.\ W., \&
Recchi, S.\ 2008, \mnras, 391, 825

\bibitem[Gaia Collaboration(2020)]{gaiaedr3}
Gaia Collaboration, Brown, A.~G.~A., Vallenari, A., et al.\ 2021, \aap, 649, 1

\bibitem[Girardi et al.(2002)]{girardi02}
Girardi, L., Bertelli, G., Bressan, A.\ et al.\ 2002, \aap, 391, 195

\bibitem[King et al.(1998)]{king98}
King, J.~R., Stephens, A., Boesgaard, M., \& Deliyannis, C.~P.\ 1998, \aj, 115, 666

\bibitem[Kurucz(2011)]{kurucz11} Kurucz, R.~L.\ 2011, Can. J. Phys., 89, 417

\bibitem[Langer et al.(1998)]{langer98}
Langer, G.~E., Fisher, D., Sneden, C., \& Bolte, M.\ 1998, \aj, 115, 685

\bibitem[Lardo et al.(2022)]{lardo22}
Lardo, C., Salaris, M., Cassisi, S., \& Bastian, N.\ 2022, \aap, 662, A117

\bibitem[Lee(2010)]{lee10} Lee, J.-W.\ 2010, \mnras, 405, L36

\bibitem[Lee(2015)]{lee15} Lee, J.-W.\ 2015, \apjs, 219, 7

\bibitem[Lee(2017)]{lee17} Lee, J.-W.\ 2017, \apj, 844, 77

\bibitem[Lee(2018)]{lee18} Lee, J.-W.\ 2018, \apjs, 238, 24

\bibitem[Lee(2019)]{lee19} Lee, J.-W.\ 2019, \apj, 872, 41

\bibitem[Lee(2021)]{lee21b} Lee, J.-W.\ 2021, \apjl, 918, L24

\bibitem[Lee(2022)]{lee22} Lee, J.-W.\ 2022, \apjs, 263, 20

\bibitem[Lee et al.(2009)]{lee09}
Lee, J.-W., Kang, Y.-W., Lee, J., \& Lee, Y.-W.\ 2009, \nat, 462, 480

\bibitem[Lee \& Sneden(2021)]{lee21a} Lee, J.-W., \& Sneden, C.\ 2021, \apj, 909, 167

\bibitem[Legnardi et al.(2022)]{legnardi22}
Legnardi, M.~V., Milone, A.~P., Armillotta, L., et al.\ 2022, \mnras, 513, 735

\bibitem[Martin et al.(2022)]{martin22}
Martin, N.~F., Ibata, R.~A., Starkenburg, E., et al.\ 2022, \mnras, 516, 5331

\bibitem[Milone et al.(2017)]{milone17}
Milone A.~P., Piotto, G., Renzini, A.\ et al.\ 2017, \mnras, 464, 3636

\bibitem[Milone et al.(2018)]{milone18}
Milone A.~P., Piotto, G., Renzini, A.\ et al.\ 2018, \mnras, 481, 5098

\bibitem[Milone et al.(2020)]{milone20}
Milone A.~P., Marino, A.~F., Da Costa, G.~S.\ et al.\ 2020, \mnras, 491, 515

\bibitem[Nardiello et al.(2018))]{uvlegacy}
Nardiello, D., Libralato, M., Piotto, G., et al.\ 2018, \mnras, 481, 3382

\bibitem[Pietrinferni et al.(2021)]{basti21}
Pietrinferni, A., Hidalgo, S., Cassisi, S., Salaris, M., et al.\ 2021, \apj, 908, 102

\bibitem[Roederer \& Sneden(2011)]{roederer11}
Roederer, I.~U., \& Sneden, C.\ 2011, \aj, 142, 22

\bibitem[Smolinski et al.(2011)]{smolinski11}
Smolinski, J.~P., Martell, S., Beers, T.~C., \& Lee, Y.~S.\ 2011, \aj, 142, 126

\bibitem[Sneden(1973)]{moog}
Sneden, C.\ 1973, PhD thesis, The University of Texas

\bibitem[Sneden(1974)]{sneden74}
Sneden, C.\ 1974, \apj, 189, 493

\bibitem[Sneden, Pilachowski, \& Kraft(2000)]{sneden00}
Sneden, C., Pilachowski, C.~A., Kraft, R.~P.\ 2000, \aj, 120, 1351

\bibitem[Sollima(2020)]{sollima20}
Sollima, A., 2020, \mnras, 495, 2222

\bibitem[Sobeck et al.(2011)]{moogscat}
Sobeck, J.~E., Kraft, R.~P., Sneden, C., et al.\ 2011, \aj, 141, 175

\bibitem[Stetson(1987)]{pbs87} 
Stetson P.~B.\ 1987, \pasp, 99, 191

\bibitem[Stetson(1994)]{pbs94} 
Stetson P.~B.\ 1994, \pasp, 106, 250

\bibitem[Suntzeff(1980)]{suntzeff80}
Suntzeff, N.~B.\ 1980, \aj, 85, 480

\bibitem[Thomas et al.(2020)]{thomas20}
Thomas, G.~F., Jensen, J., McConnachie, A., et al.\ 2020, \aj, 902, 89

\bibitem[Thurl \& Johnston(2002)]{thurl02}
Thurl, C., \& Johnston, K.~V.\ 2002, ASP Conf.\ Ser.\ Vol.\ 265,
$\omega$ Centauri: A Unique Window into Astrophysics,
ed.\ F.\ van Leeuwen, J.~D.\ Hughes, \& G.\ Piotto (San Francisco: ASP), 337

\end{thebibliography}
\end{document}